\documentclass[aps,prl,reprint,superscriptaddress,amsmath,amssymb,floatfix,nofootinbib,nobibnotes]{revtex4-2}

\usepackage{amsmath}
\usepackage{amssymb}
\usepackage{graphicx}
\graphicspath{{./}}

\usepackage{color}

\date{}

\begin{document}

\title{Structure Selection by Non-Conservative 3-Body Acoustic Interactions}

\author{Qinghao Mao}
\email{qinghaomao@uchicago.edu}
\author{Heinrich M. Jaeger}
\affiliation{
\normalsize{Department of Physics, University of Chicago,}\\
\normalsize{929 E 57th St, Chicago, IL 60637, USA}\\
}
\affiliation{
\normalsize{James Franck Institute, University of Chicago,}\\
\normalsize{929 E 57th St, Chicago, IL 60637, USA}\\
}

\date{\today}

\begin{abstract}
Non-conservative multi-body interactions are typically associated with instabilities and activity in driven, field-mediated systems. Here we show that they can also promote stable static structures. Combining experiments and simulations in a minimal, acoustically levitated three-particle system, we tune the relative strength of conservative and non-conservative contributions to the force field. The conservative component favors a symmetric equilibrium configuration, whereas the non-conservative 3-body contribution selects a flattened isosceles triangle. Our results identify non-conservative multi-body forces as a mechanism for static structure selection in driven-dissipative matter in the absence of an effective-energy landscape.
\end{abstract}

\maketitle

\textit{Introduction—}The standard starting point for predicting the structure of particle ensembles is the minimization of an effective interaction energy \cite{israelachvili2011BOOK,russel1989BOOK}. However, in many driven-dissipative systems, interactions intrinsically contain non-conservative components that cannot be derived from an energy \cite{fruchart2021NATURE,du2025ARXIV,king2025PRR,morrell2026PRL}. Furthermore, such interactions often cannot be expressed as pairwise additive, since the force on each particle depends on the full configuration of the ensemble and must therefore be treated as multi-body \cite{zampetaki2021PNAS,dobnikar2003JCP,dobnikar2004PRE,merrill2009PRL}. Examples are found in active matter and in particle ensembles with interactions mediated by acoustic, optical, hydrodynamic, or chemical fields \cite{lim2024RPP,stclair2023PRR,wu2025PRR,davenport2022SOFTMATTER,forbes2020NANOPHOTONICS,huang2022NATCOMM,parker2025NATCOMM}. Generally, a non-conservative multi-body force component is expected to drive a system away from the stable equilibrium structure associated with the conservative part. Past work has shown that this can introduce spontaneous dynamics in ensembles of individually passive particles \cite{haefner2009PRL,wu2009PRL,li2021NATCOMM,parker2025NATCOMM, shi2025PNAS}, including self-propulsion and limit-cycle behavior \cite{wu2025PRE}. Here we ask whether it can also play the opposite role: rather than merely driving motion, can non-conservative multi-body interactions select stable static structures?

In this Letter, we combine experiments and simulations to test this question in a minimal three-particle system levitated in an acoustic trap in which we can tune the relative strength of conservative and non-conservative contributions to the multi-body interactions. We demonstrate that, while the conservative component favors a symmetric equilibrium structure, increasing the relative strength of the non-conservative 3-body interaction introduces a bifurcation that drives the system toward lower-symmetry configurations. In our system, this mechanism selects a flattened isosceles triangle that is force-balanced and stable, but distinct from the structure predicted by the conservative interaction energy alone.

\begin{figure}[t]
    \centering
    \includegraphics[width=72mm]{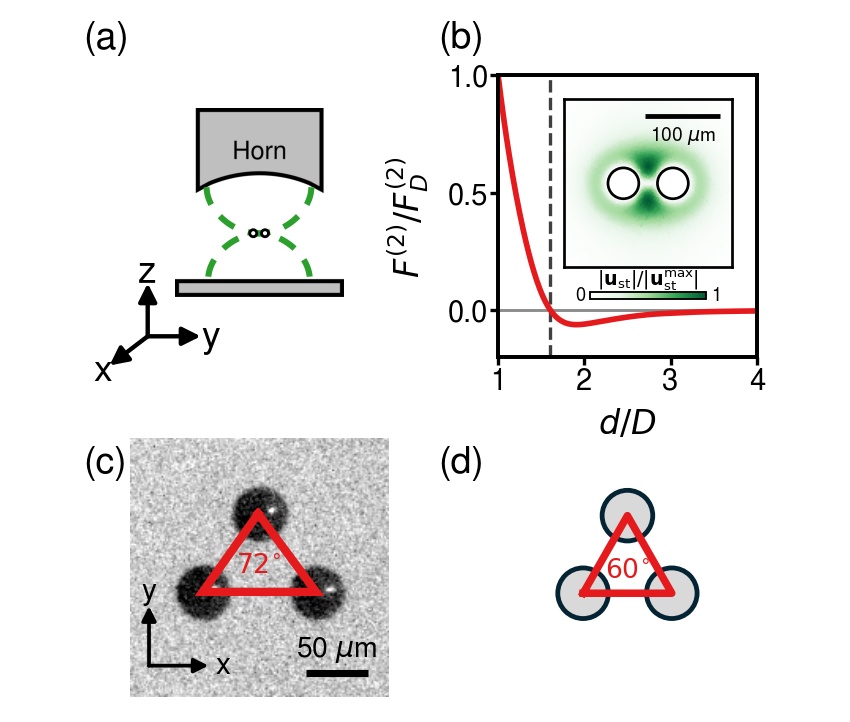}
    \caption{Experimental observation of symmetry breaking in an ensemble of three acoustically levitated particles. (a) Sketch of  the experimental acoustic levitation geometry, with the standing sound wave indicated between an ultrasound transmitter (horn) and a reflecting plate. (b) Force between particles in an isolated pair, normalized by $F^{(2)}_D$, versus center-to-center distance $d/D$, where $F^{(2)}_D$ is the force when two particles are in contact; the dashed line marks the stable pair spacing used in the additive pairwise model. Inset: two-particle microstreaming field, where $u_{st}$ is the local streaming velocity shown in green. (c) Example of an experimentally observed three-particle configuration, here with a maximum internal angle of about $72^\circ$. (d) For three particles with pairwise additive interactions, because all three pair forces are balanced by symmetry, the predicted configuration is an equilateral triangle.}
    \label{fig:observation}
\end{figure}

\textit{Experiments—}Fig.~\ref{fig:observation}(a) sketches the levitation setup used in the experiment. The particle system consists of three identical polystyrene spheres with diameter $D=(41.1\pm0.5)\,\mu\mathrm{m}$, confined to a two-dimensional levitation plane  in the center of an acoustic trap formed by a single-axis Langevin-type ultrasonic transducer with an aluminum horn \cite{wu2023PNAS,lim2024RPP,wu2025PRE,mao2025NATCOMM}.

The particles interact through the surrounding acoustic field, and the force on an isolated pair of particles is shown in Fig.~\ref{fig:observation}(b). This pairwise force arises from the balance between a long-range attraction due to scattered sound and a short-range repulsion due to sound-induced micro streaming, producing a stable center-to-center distance\cite{fabre2017JFM, wu2023PNAS, wu2025PRE}. For three particles with isotropic, pairwise additive interactions, the stable configuration then is an equilateral triangle: all three pair separations are identical and force-balanced. In our experiments, however, the observed stable three-particle structure is not equilateral. An example is shown in Fig.~\ref{fig:observation}(c), where the particles form a flattened isosceles triangle with a maximum internal angle of about $72^\circ$, in contrast to the $60^\circ$ predicted by the pairwise model, Fig.~\ref{fig:observation}(d).

This non-equilateral isosceles configuration is not imposed by a fixed laboratory anisotropy. Across repeated trials, the flattened configuration appears in all of the three particle-permutation-related orientations rather than along one preferred direction in the imaging plane. The measured angular splitting exceeds the tracking uncertainty and persists when particles are exchanged, and the data show no corresponding preferred direction. The frame-by-frame largest-angle and orientation distributions for a representative three-particle run are shown in Fig.~\ref{fig:supp-2026052201-histograms}. These checks distinguish the observed symmetry breaking from trap anisotropy, particle mismatch, or image-analysis bias.

This comparison reveals the key puzzle: the experimentally observed three-particle structure cannot be predicted by summing isotropic pair forces and, therefore, the observed non-equilateral state signals a force contribution that exists only with a simultaneous three-particle configuration.

\begin{figure}[t]
    \centering
    \includegraphics[width=0.5\textwidth]{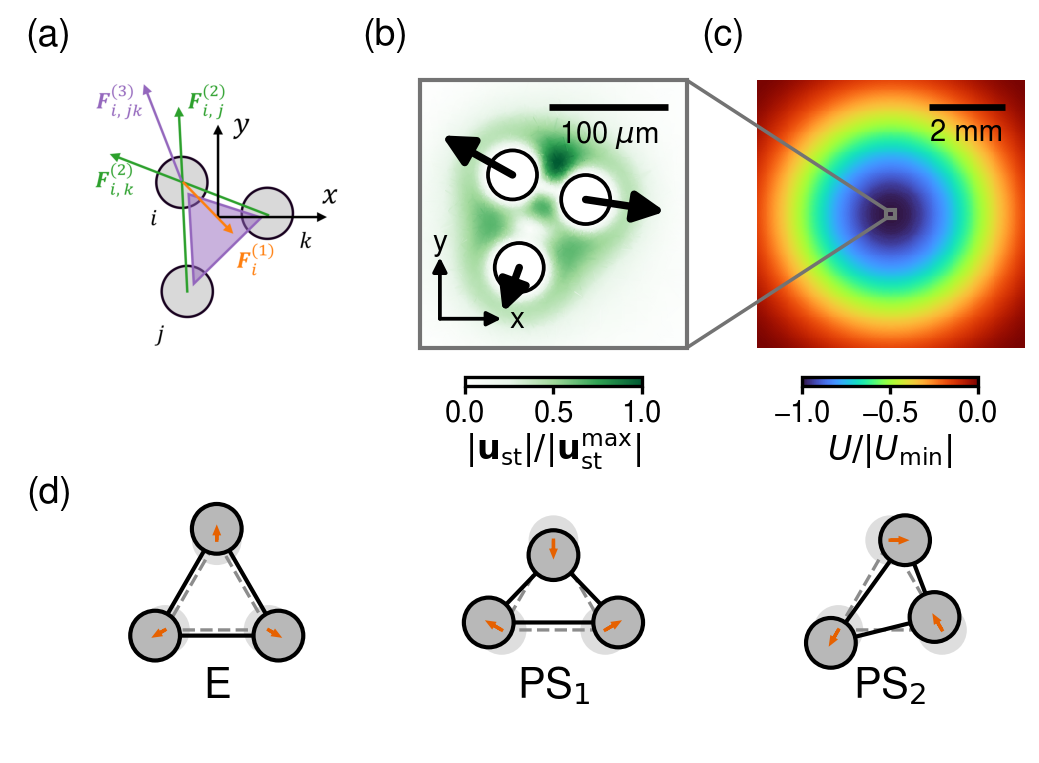}
    \caption{Forces acting on the particles in the levitation plane. (a) Local steady-state microstreaming velocity field magnitude $|{\bf u}_\text{st}|$ around a three-particle configuration, with arrows indicating the induced motion directions. (b) Lateral acoustic trapping potential $U$ in the x-y levitation plane, giving rise to the 1-body confinement.  (c) Vector decomposition of the force on particle $i$ into 2-body terms $\mathbf{F}_{i,j}^{(2)}$ and $\mathbf{F}_{i,k}^{(2)}$ and the additional 3-body term $\mathbf{F}_{i,jk}^{(3)}$. (d) Isotropic expansion mode $E$ and pure-shear modes $\mathrm{PS}_{1}$ and $\mathrm{PS}_{2}$ used for the deformation-space projection.}
    \label{fig:force_decomp}
\end{figure}

\textit{1-body, 2-body, and 3-body forces—}Figure~\ref{fig:force_decomp} shows the force-decomposition procedure used to identify the origin of the symmetry-broken structure. We write the total force on particle $i$ as a sum of a 1-body term, two isolated pair terms, and a residual 3-body term:
\begin{equation}
    \mathbf{F}_{i}
    =
    \mathbf{F}_{i}^{(1)}
    +
    \mathbf{F}_{i,j}^{(2)}
    +
    \mathbf{F}_{i,k}^{(2)}
    +
    \mathbf{F}_{i,jk}^{(3)}.
    \label{eq:force-decomp}
\end{equation}

The 2-body contribution is constructed by asking how particle $i$ would be forced by isolated interactions with particles $j$ and $k$. In the vector construction in Fig.~\ref{fig:force_decomp}(a), this pairwise part is written as
$\mathbf{F}_{i}^{(2)} = \mathbf{F}_{i,j}^{(2)} + \mathbf{F}_{i,k}^{(2)}.$
In this construction, each pair interaction is independent of the third particle.

The additional 3-body contribution is the part of the force on particle $i$ that requires the simultaneous three-particle configuration, as indicated by the local streaming field in Fig.~\ref{fig:force_decomp}(b), $\mathbf{F}_{i}^{(3)} = \mathbf{F}_{i,jk}^{(3)}$.
Physically, this term arises because scattering and microstreaming around the particles are determined by the full three-particle geometry, not by independent pairs.

We also include a 1-body term that arises from the shape of the acoustic trapping potential within the $x$-$y$ plane. This contribution is obtained from the acoustic confinement landscape in Fig.~\ref{fig:force_decomp}(c). In the levitation plane, the Gor'kov potential generates a force that pulls particles toward the center \cite{gorkov1962SOVPHYS,bruus2012LABCHIP,settnes2012PRE,karlsen2015PRE}. It does not depend on the positions of any other particles.

\textit{Shape-space description—}To analyze how the triangle changes shape, we project the particle displacements onto the collective modes shown in Fig.~\ref{fig:force_decomp}(d). Three particles in two dimensions have six in-plane degrees of freedom. Three of these correspond to rigid-body motion: translation in $x$ ($T_x$), translation in $y$ ($T_y$), and global rotation ($R$), which do not change the shape of the triangle, while the remaining three modes change the internal geometry. The isotropic expansion mode $E$ expands or contracts the triangle uniformly. The two pure-shear modes $\mathrm{PS}_{1}$ and $\mathrm{PS}_{2}$ deform an equilateral triangle into an isosceles or scalene shape. Because we focus on shape change, the deformation coordinates
\begin{equation}
    \Delta = \left(\mathrm{E},\mathrm{PS}_{1},\mathrm{PS}_{2}\right)
\end{equation}
provide a natural coordinate system for studying it. The projected deformation force then is written as
\begin{equation}
    \mathbf{F}_{\Delta}^{\mathrm{def}}
    =
    F_{\Delta}^\mathrm{E}\mathbf{e}_\mathrm{E}
    +
    F_{\Delta}^{\mathrm{PS}_{1}}\mathbf{e}_{\mathrm{PS}_{1}}
    +
    F_{\Delta}^{\mathrm{PS}_{2}}\mathbf{e}_{\mathrm{PS}_{2}} .
    \label{eq:deformation-force}
\end{equation}

In this representation, the equilateral configuration sits at the origin of the $\mathrm{PS}_{1}$-$\mathrm{PS}_{2}$ plane. A stable equilateral triangle requires the net force to point back toward this origin for small shape perturbations. If instead the force points outward along a pure-shear direction, the equilateral state becomes unstable and the system moves toward a symmetry-broken configuration.

\begin{figure}[t]
    \centering
\includegraphics[width=\columnwidth]{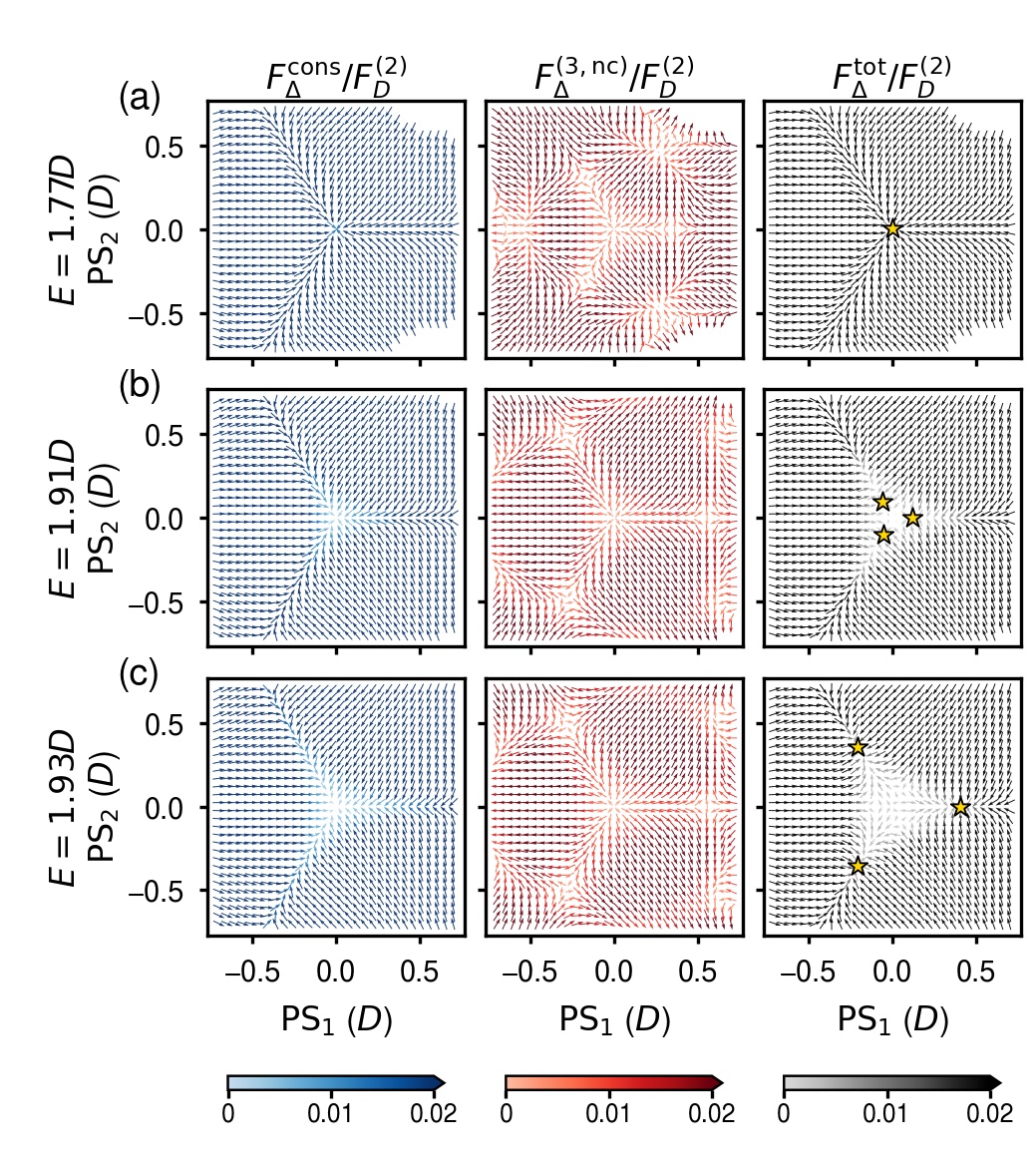}
    \caption{Force fields in the $\mathrm{PS}_{1}$-$\mathrm{PS}_{2}$ pure-shear plane for representative values of $E$. The equilateral side length $E$ and deformation coordinates $\mathrm{PS}_{1}$ and $\mathrm{PS}_{2}$ are reported in units of the particle diameter $D$. Colors show the force magnitude normalized by $F_{D}^{(2)}$. Columns show the combined conservative contribution $\mathbf{F}_{\Delta}^{\mathrm{cons}}=\mathbf{F}_{\Delta}^{(1)}+\mathbf{F}_{\Delta}^{(2)}+\mathbf{F}_{\Delta}^{(3,\mathrm{cons})}$, the non-conservative 3-body contribution $\mathbf{F}_{\Delta}^{(3,\mathrm{nc})}$, and the total force $\mathbf{F}_{\Delta}^{\mathrm{tot}}$. Here $\mathrm{nc}$ denotes non-conservative. Stars mark stable roots of the total projected force.}
    \label{fig:shape_space}
\end{figure}

\textit{Force decomposition in shape space—}Figure~\ref{fig:shape_space} shows how the competing force contributions act in the shape-space coordinates. In our system, the 1-body confinement and 2-body pair forces are conservative; after projection onto the three-dimensional deformation space, the residual 3-body force is delineated into conservative and non-conservative components using a Helmholtz decomposition,
\begin{equation}
    \mathbf{F}_{\Delta}^{(3)}
    =
    \mathbf{F}_{\Delta}^{(3,\mathrm{cons})}
    +
    \mathbf{F}_{\Delta}^{(3,\mathrm{nc})}
    =
    \nabla_{\Delta}\phi
    +
    \nabla_{\Delta}\times\mathbf{A}.
    \label{eq:helmholtz}
\end{equation}
Here $\phi$ is a scalar potential in the deformation space and $\mathbf{A}$ is a vector potential. The numerical procedure is described in the End Matter. The three columns of Fig.~\ref{fig:shape_space} show, for different  triangle configurations at three values of isotropic expansion $\mathrm{E}$,  the fields representing the sum of all conservative forces, the non-conservative component of the 3-body force, and the total force.
The conservative force restores the equilateral geometry, whereas the non-conservative 3-body force has the sign required to destabilize the equilateral pure-shear modes \cite{fruchart2021NATURE,du2025ARXIV}. In Fig.~\ref{fig:shape_space}(a), the restoring forces dominate and the only stable structure is the equilateral triangle. In Fig.~\ref{fig:shape_space}(b) and Fig.~\ref{fig:shape_space}(c), the equilateral state loses stability and the total force develops three symmetry-related stable states. Each state corresponds to a flattened triangle, with one of the three particles occupying the distinct vertex of the isosceles configuration. We note  that the flattened triangle should self-propel if the 1-body confining force can be ignored \cite{wu2025PRE}; however, here the 1-body force counteracts this tendency, and the trimer is fully force-balanced and therefore static.

\begin{figure*}[t]
    \centering
    \includegraphics[width=\textwidth]{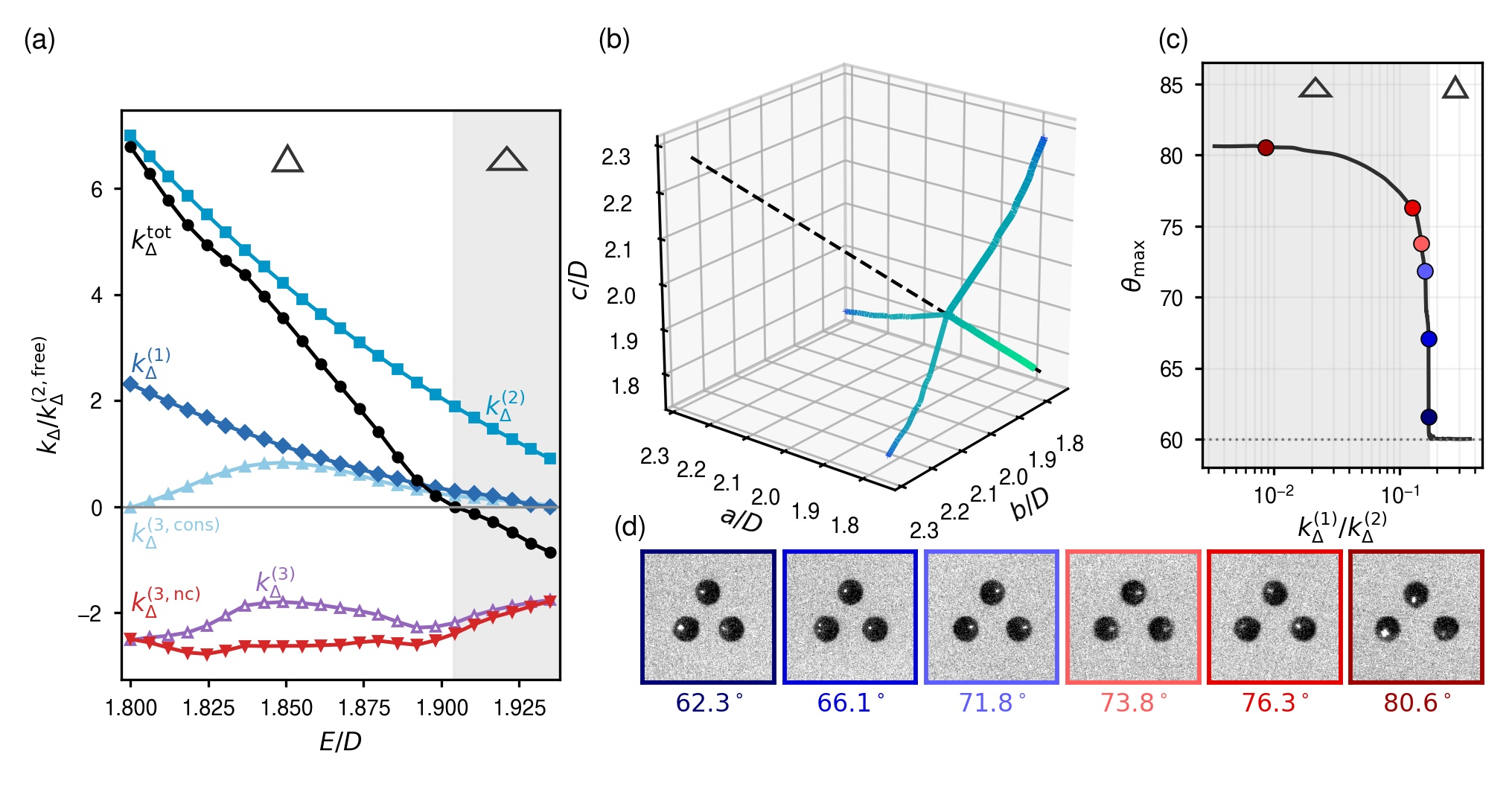}
    \caption{Tuning the non-conservative structure selection. (a) Pure-shear restoring stiffness contributions versus equilateral expansion $E$, normalized by the confinement-free pairwise equilateral restoring stiffness $k_{\Delta}^{(2,\mathrm{free})}$; negative total stiffness marks the bifurcated isosceles regime. (b) Ten-million-step MD relaxation trajectory in side-length space $(a,b,c)$ relative to the equilateral dashed line. (c) Maximum internal angle versus $k_{\Delta}^{(1)}/k_{\Delta}^{(2)}$, with colored markers indicating the experimental image sequence. (d) Raw experimental images showing the evolution from a nearly equilateral triangle to a flattened isosceles state.}
    \label{fig:tuning}
\end{figure*}

\textit{Tuning the structure—}The force field retains the permutation symmetry of three identical particles, but the stable fixed point does not: the system selects one of the three symmetry-related flattened states. To see this finite-cluster symmetry breaking and structure selection more quantitatively, we analyze the restoring stiffness of the equilateral state. Fig.~\ref{fig:tuning}(a) quantifies the competition between stabilizing and destabilizing forces by plotting the center restoring stiffness along a pure-shear direction as a function of the expansion coordinate $E$. In this sign convention, positive total stiffness corresponds to a stable equilateral state, whereas negative total stiffness means that the equilateral state is unstable to pure-shear deformation and a new, stable flattened structure will take its place. The non-conservative 3-body force is the destabilizing contribution; as the conservative restoring forces become weaker at larger $E$, the balance shifts and the total stiffness reverses sign. This produces a bifurcation in the trimer shape space: the symmetric fixed point loses stability and three symmetry-related flattened fixed points appear. It resembles a continuous symmetry-breaking transition in the sense that an order parameter grows from zero, but no equilibrium free energy is defined \cite{fruchart2021NATURE,du2025ARXIV}.

We can see this bifurcation more clearly in Fig.~\ref{fig:tuning}(b). Let $a$, $b$, and $c$ denote the three edge lengths of the triangle. Equilateral configurations lie along the line $a=b=c=\mathrm{E}$. When the restoring forces dominate, the stable state remains on this equilateral line. When the non-conservative 3-body contribution dominates, the stable state moves away from the line, showing that the triangle has entered a symmetry-broken, but still stable, configuration.

Fig.~\ref{fig:tuning}(c) reduces this structural change to a single observable: the maximum internal angle. As confinement is reduced, the maximum internal angle increases sharply from $60^\circ$ to approximately $81^\circ$, marking the opening of the triangle. This angle provides a direct measurable signature of which static structure the system has selected in experiment.

\textit{Experimental control of structure—}The stability analysis in Fig.~\ref{fig:tuning}(a-c) shows that the trimer shape is set by the balance between conservative restoring forces and the non-conservative 3-body destabilizing force. Experimentally, we tune this balance by changing the acoustic power, see Fig.~\ref{fig:supp-gorkov-slices}. Reducing acoustic power decreases the levitation height and weakens the effective in-plane confinement, moving the system across the stability boundary predicted by the force decomposition.

The raw experimental images in Fig.~\ref{fig:tuning}(d) show the corresponding change in triangle geometry. At stronger confinement, the three particles remain close to an equilateral triangle. As confinement weakens, the trimer opens and becomes increasingly flattened, with the maximum internal angle growing from $60^\circ$ toward approximately $81^\circ$. Although the confinement-strength distribution inside the cavity is difficult to measure directly, the simulated angle curve in Fig.~\ref{fig:tuning}(c) reproduces the qualitative experimental trend, showing that the observed shape change is consistent with a confinement-controlled stability transition.

\textit{Conclusion—}Our results demonstrate that non-conservative multi-body interactions can do more than destabilize particle ensembles or generate spontaneous motion. In the acoustically levitated trimer studied here, the non-conservative 3-body contribution selects a force-balanced and linearly stable isosceles configuration that is not predicted when only the conservative interaction energies are considered. This minimal system provides a prototypical example for which the analysis of the competition between conservative and non-conservative force components can be carried out straightforwardly. However, the same ideas should apply also to ensembles of more than three identical particles whose configuration is selected by non-conservative multi-body forces \cite{mao2025NATCOMM, zampetaki2021PNAS}. More broadly, we expect that this mechanism is quite general and should be relevant in a wide range of driven-dissipative systems with field-mediated interactions, including also optical, hydrodynamic, chemical, and active-matter settings, where non-conservative force components may be used not only to drive motion, but to program identical passive objects into stable structures beyond those accessible through an energy landscape.

\textit{Acknowledgements—}We thank Nina Brown, Danny Seara, Brady Wu, and Robert Hunt for helpful discussions. This work was supported by the NSF through DMR-2104733. The
experiments utilized the shared experimental facilities at the
University of Chicago MRSEC, which is funded by the
National Science Foundation under award number DMR-2011854. The research also benefited from computational
resources and services supported by the Research Computing Center at the University of Chicago.

\bibliography{ref}

@article{king2025PRR,
  title={Scattered waves fuel emergent activity},
  author={King, Ella M and Morrell, Mia C and Sustiel, Jacqueline B and Gronert, Matthew and Pastor, Hayden and Grier, David G},
  journal={Physical Review Research},
  volume={7},
  number={1},
  pages={013055},
  year={2025},
  publisher={APS}
}

@article{morrell2026PRL,
  title={Nonreciprocal wave-mediated interactions power a classical time crystal},
  author={Morrell, Mia C and Elliott, Leela and Grier, David G},
  journal={Physical Review Letters},
  volume={136},
  number={5},
  pages={057201},
  year={2026},
  publisher={APS}
}

@article{shi2025PNAS,
  title={Electrostatics overcome acoustic collapse to assemble, adapt, and activate levitated matter},
  author={Shi, Sue and H{\"u}bl, Maximilian C and Grosjean, Galien and Goodrich, Carl P and Waitukaitis, Scott},
  journal={Proceedings of the National Academy of Sciences},
  volume={122},
  number={50},
  pages={e2516865122},
  year={2025},
  publisher={National Academy of Sciences}
}

@article{davenport2022SOFTMATTER,
  title = {Formation of Colloidal Chains and Driven Clusters with Optical Binding},
  author = {Davenport, Dominique J. and Kleckner, Dustin},
  year = {2022},
  journal = {Soft Matter},
  shortjournal = {Soft Matter},
  volume = {18},
  number = {23},
  pages = {4464--4474},
  publisher = {The Royal Society of Chemistry},
  issn = {1744-6848},




  langid = {english},

}

@article{dobnikar2003JCP,
  title = {Many-Body Interactions and the Melting of Colloidal Crystals},
  author = {Dobnikar, J. and Chen, Y. and Rzehak, R. and von Grünberg, H. H.},
  year = {2003},
  journal = {The Journal of Chemical Physics},
  shortjournal = {J. Chem. Phys.},
  volume = {119},
  number = {9},
  pages = {4971--4985},
  issn = {0021-9606},



}

@article{dobnikar2004PRE,
  title = {Three-Body Interactions in Colloidal Systems},
  author = {Dobnikar, Jure and Brunner, Matthias and Von Grünberg, Hans-Hennig and Bechinger, Clemens},
  year = {2004},
  journal = {Physical Review E},
  shortjournal = {Phys. Rev. E},
  volume = {69},
  number = {3},
  pages = {031402},
  issn = {1539-3755, 1550-2376},

  langid = {english},

}

@article{du2025ARXIV,
  title = {Hidden Nonreciprocity as a Stabilizing Effective Potential in Active Matter},
  author = {Du, Matthew and Goychuk, Andriy and Vaikuntanathan, Suriyanarayanan},
  year = {2024},
  eprint = {2401.14690},
  journal = {arXiv},
  eprintclass = {cond-mat.stat-mech},


  pubstate = {prepublished},

}

@article{fruchart2021NATURE,
  title = {Non-Reciprocal Phase Transitions},
  author = {Fruchart, Michel and Hanai, Ryo and Littlewood, Peter B. and Vitelli, Vincenzo},
  year = {2021},
  journal = {Nature},
  volume = {592},
  number = {7854},
  pages = {363--369},
  publisher = {Nature Publishing Group},
  issn = {1476-4687},




  langid = {english},

}

@article{huang2022NATCOMM,
  title = {The Primeval Optical Evolving Matter by Optical Binding inside and Outside the Photon Beam},
  author = {Huang, Chih-Hao and Louis, Boris and Bresolí-Obach, Roger and Kudo, Tetsuhiro and Camacho, Rafael and Scheblykin, Ivan G. and Sugiyama, Teruki and Hofkens, Johan and Masuhara, Hiroshi},
  year = {2022},
  journal = {Nature Communications},
  shortjournal = {Nat Commun},
  volume = {13},
  number = {1},
  pages = {5325},
  publisher = {Nature Publishing Group},
  issn = {2041-1723},




  langid = {english},

}

@book{israelachvili2011BOOK,
  title = {Intermolecular and {{Surface Forces}}},
  author = {Israelachvili, Jacob N.},
  year = {2011},
  publisher = {Academic Press},

  isbn = {978-0-12-391933-5},
  langid = {english},
  pagetotal = {708}
}

@article{merrill2009PRL,
  title = {Many-{{Body Electrostatic Forces}} between {{Colloidal Particles}} at {{Vanishing Ionic Strength}}},
  author = {Merrill, Jason W. and Sainis, Sunil K. and Dufresne, Eric R.},
  year = {2009},
  journal = {Physical Review Letters},
  shortjournal = {Phys. Rev. Lett.},
  volume = {103},
  number = {13},
  pages = {138301},
  doi = {10.1103/PhysRevLett.103.138301},


}

@article{parker2025NATCOMM,
  title = {Symmetry Breaking-Induced {{N-body}} Electrodynamic Forces in Optical Matter Systems},
  author = {Parker, John and Nagasamudram, Spoorthi and Peterson, Curtis and Li, Yanzeng and Soleimanikahnoj, Sina and Rice, Stuart A. and Scherer, Norbert F.},
  year = {2025},
  journal = {Nature Communications},
  shortjournal = {Nat Commun},
  volume = {16},
  number = {1},
  pages = {6294},
  publisher = {Nature Publishing Group},
  issn = {2041-1723},




  langid = {english},

}

@book{russel1989BOOK,
  title = {Colloidal {{Dispersions}}},
  author = {Russel, W. B. and Saville, D. A. and Schowalter, W. R.},
  year = {1989},
  publisher = {Cambridge University Press},
  location = {Cambridge},




  isbn = {978-0-521-42600-8},

}

@article{wu2023PNAS,
  title = {Hydrodynamic Coupling Melts Acoustically Levitated Crystalline Rafts},
  author = {Wu, Brady and VanSaders, Bryan and Lim, Melody X. and Jaeger, Heinrich M.},
  year = {2023},
  journal = {Proceedings of the National Academy of Sciences},
  shortjournal = {Proc. Natl. Acad. Sci. U.S.A.},
  volume = {120},
  number = {29},
  pages = {e2301625120},
  issn = {0027-8424, 1091-6490},




  langid = {english},

}

@article{wu2025PRE,
  title = {Nonreciprocity and Multibody Interactions in Acoustically Levitated Particle Systems: {{A}} Three-Body Problem},
  shorttitle = {Nonreciprocity and Multibody Interactions in Acoustically Levitated Particle Systems},
  author = {Wu, Brady and Mao, Qinghao and VanSaders, Bryan and Jaeger, Heinrich M.},
  year = {2025},
  journal = {Physical Review E},
  shortjournal = {Phys. Rev. E},
  volume = {112},
  number = {3},
  pages = {035410},
  publisher = {American Physical Society},





}

@article{mao2025NATCOMM,
  title = {Structural Reconfiguration of Interacting Multi-Particle Systems through Parametric Pumping},
  author = {Mao, Qinghao and Wu, Brady and VanSaders, Bryan and Jaeger, Heinrich M.},
  year = {2025},
  journal = {Nature Communications},
  shortjournal = {Nat. Commun.},
  volume = {16},
  pages = {4637},
  doi = {10.1038/s41467-025-59631-3}
}

@article{zampetaki2021PNAS,
  title = {Collective Self-Optimization of Communicating Active Particles},
  author = {Zampetaki, Alexandra V. and Liebchen, Benno and Ivlev, Alexei V. and Löwen, Hartmut},
  year = {2021},
  journal = {Proceedings of the National Academy of Sciences},
  volume = {118},
  number = {49},
  pages = {e2111142118},
  publisher = {Proceedings of the National Academy of Sciences},





}

@article{gorkov1962SOVPHYS,
  title = {On the Forces Acting on a Small Particle in an Acoustical Field in an Ideal Fluid},
  author = {Gor'kov, L. P.},
  year = {1962},
  journal = {Soviet Physics Doklady},
  volume = {6},
  pages = {773--775}
}

@article{bruus2012LABCHIP,
  title = {Acoustofluidics 7: The Acoustic Radiation Force on Small Particles},
  author = {Bruus, Henrik},
  year = {2012},
  journal = {Lab on a Chip},
  shortjournal = {Lab Chip},
  volume = {12},
  number = {6},
  pages = {1014--1021},

}

@article{settnes2012PRE,
  title = {Forces Acting on a Small Particle in an Acoustical Field in a Viscous Fluid},
  author = {Settnes, Mikkel and Bruus, Henrik},
  year = {2012},
  journal = {Physical Review E},
  shortjournal = {Phys. Rev. E},
  volume = {85},
  number = {1},
  pages = {016327},

}

@article{karlsen2015PRE,
  title = {Forces Acting on a Small Particle in an Acoustical Field in a Thermoviscous Fluid},
  author = {Karlsen, Jonas T. and Bruus, Henrik},
  year = {2015},
  journal = {Physical Review E},
  shortjournal = {Phys. Rev. E},
  volume = {92},
  number = {4},
  pages = {043010},

}

@article{stclair2023PRR,
  title = {Dynamics of Acoustically Bound Particles},
  author = {St. Clair, Nicholas and Davenport, Dominique and Kim, Arnold D. and Kleckner, Dustin},
  year = {2023},
  journal = {Physical Review Research},
  shortjournal = {Phys. Rev. Res.},
  volume = {5},
  number = {1},
  pages = {013051},

}

@article{lim2024RPP,
  title = {Acoustic Manipulation of Multi-Body Structures and Dynamics},
  author = {Lim, Melody X. and VanSaders, Bryan and Jaeger, Heinrich M.},
  year = {2024},
  journal = {Reports on Progress in Physics},
  shortjournal = {Rep. Prog. Phys.},
  volume = {87},
  number = {6},
  pages = {064601},

}

@article{wu2025PRR,
  title = {Pattern Formation in Acoustically Levitated Particle Systems with Competing Near-Field Interactions},
  author = {Wu, Brady and Esposito, Edward P. and Mao, Qinghao and Jaeger, Heinrich M.},
  year = {2025},
  journal = {Physical Review Research},
  shortjournal = {Phys. Rev. Res.},
  volume = {7},
  number = {2},
  pages = {023017},

}

@article{haefner2009PRL,
  title = {Conservative and Nonconservative Torques in Optical Binding},
  author = {Haefner, David and Sukhov, Sergey and Dogariu, Aristide},
  year = {2009},
  journal = {Physical Review Letters},
  shortjournal = {Phys. Rev. Lett.},
  volume = {103},
  number = {17},
  pages = {173602},

}

@article{wu2009PRL,
  title = {Direct Measurement of the Nonconservative Force Field Generated by Optical Tweezers},
  author = {Wu, Pinyu and Huang, Rongxin and Tischer, Christian and Jonas, Alexandr and Florin, Ernst-Ludwig},
  year = {2009},
  journal = {Physical Review Letters},
  shortjournal = {Phys. Rev. Lett.},
  volume = {103},
  number = {10},
  pages = {108101},

}

@article{li2021NATCOMM,
  title = {Non-Hermitian Physics for Optical Manipulation Uncovers Inherent Instability of Large Clusters},
  author = {Li, Xiao and Liu, Yineng and Lin, Zhifang and Ng, Jack and Chan, C. T.},
  year = {2021},
  journal = {Nature Communications},
  shortjournal = {Nat Commun},
  volume = {12},
  number = {1},
  pages = {6597},

}

@article{forbes2020NANOPHOTONICS,
  title = {Optical Binding of Nanoparticles},
  author = {Forbes, Kayn A. and Bradshaw, David S. and Andrews, David L.},
  year = {2020},
  journal = {Nanophotonics},
  volume = {9},
  number = {1},
  pages = {1--17},

}

@article{fabre2017JFM,
  title = {Acoustic Streaming and the Induced Forces between Two Spheres},
  author = {Fabre, D. and Jalal, J. and Leontini, J. S. and Manasseh, R.},
  year = {2017},
  journal = {Journal of Fluid Mechanics},
  shortjournal = {J. Fluid Mech.},
  volume = {810},
  pages = {378--391},
  issn = {0022-1120, 1469-7645},

}

\clearpage
\setcounter{equation}{0}
\renewcommand{\theequation}{S\arabic{equation}}
\setcounter{figure}{0}
\renewcommand{\thefigure}{S\arabic{figure}}
{\centering\bfseries End Matter\par}

\textit{Experiments—}The acoustic levitation apparatus follows the setup described in our previous work on acoustic particle assemblies \cite{wu2025PRE,mao2025NATCOMM}. A single-axis Langevin transducer is attached to an amplifying aluminum horn, whose lower surface forms the upper boundary of the acoustic cavity. A transparent ITO-coated glass reflector forms the lower boundary and allows bottom-view imaging. The horn-reflector spacing is adjusted to approximately one half acoustic wavelength at the horn resonance, near $34.8\mathrm{kHz}$, so that the standing sound wave levitates particles in a horizontal plane just below the pressure node. The horn surface has a curvature of $R=25\mathrm{mm}$, producing the radial gradient in the acoustic field that confines particles near the center of the levitation plane. The piezoelectric stack is driven by a function generator and high-voltage amplifier; the acoustic pressure is monitored near the cavity. Any residual charge on the particles is neutralized with a photo-ionizer before measurements. The particles are polystyrene microspheres (microParticles GmbH) with density $\rho\simeq1.05\,\mathrm{g\,cm^{-3}}$ and diameter $D=(41.1\pm0.5)\,\mu\mathrm{m}$. Particles are introduced one at a time with a fine probe, and their in-plane configurations are recorded from below with high-speed video. Fig.~\ref{fig:supp-2026052201-histograms} shows frame-by-frame statistics from a representative three-particle experiment, including the distribution of the maximum internal angle and the orientation of the maximum-angle vertex.

\begin{figure}[h]
    \centering
    \includegraphics[width=0.45\textwidth]{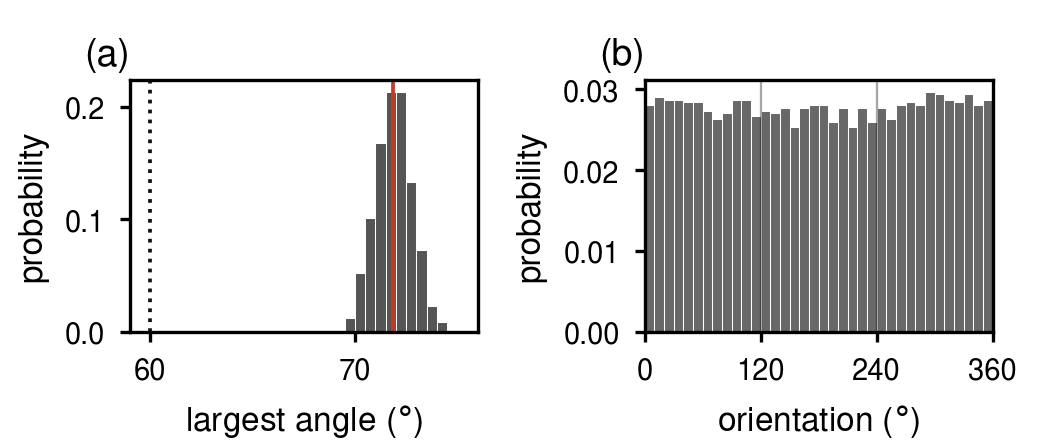}
    \caption{Frame-by-frame statistics for the low-power three-particle experiment at $34750\mathrm{Hz}$. (a) Probability distribution of the maximum internal angle. The dotted line marks the $60^\circ$ equilateral value, and the red line marks the mean angle. (b) Orientation distribution of the maximum-angle vertex.}
    \label{fig:supp-2026052201-histograms}
\end{figure}

\textit{FEM simulations—}The acoustic forces were calculated with finite-element method (FEM) simulations implemented in COMSOL Multiphysics, following the workflow used in our previous acoustic-levitation studies \cite{wu2025PRE,mao2025NATCOMM}.
Each force calculation was carried out in two steps. First, we solved the linearized acoustic problem in the frequency domain using the Thermoviscous Acoustics interface to get the first-order pressure and velocity fields, including viscous boundary-layer effects near the particle surfaces.  Second, the time-averaged second-order flow was computed with the Laminar Flow interface by using the source terms generated from the first-order acoustic solution. This gives the steady streaming flow around the particles.

\begin{figure}[t]
    \centering
    \includegraphics[width=0.45\textwidth]{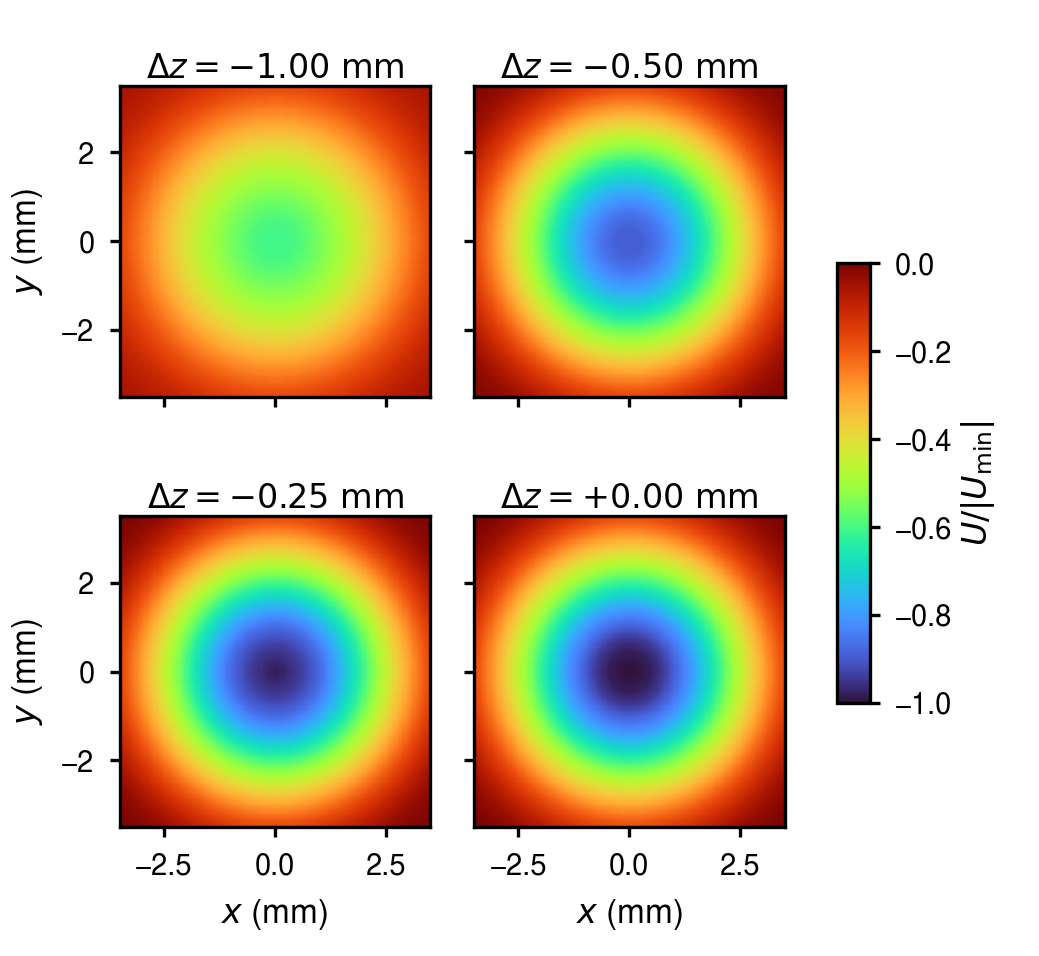}
    \caption{No-particle Gor'kov-potential cross sections from the FEM horn simulation used to define the 1-body confinement. The four panels show horizontal slices at offsets $\Delta z=-1.00,-0.50,-0.25$, and $0\,\mathrm{mm}$ relative to the Gor'kov-potential minimum for the $3.80\,\mathrm{mm}$ horn spacing. The potential is normalized by the magnitude of the minimum.}
    \label{fig:supp-gorkov-slices}
\end{figure}

For each fixed particle configuration, the force on every particle was obtained by integrating the pressure and viscous stress over the particle surface. We refer to these two contributions as the scattering and streaming parts of the acoustic force. The total in-plane force is the sum of these surface-integrated contributions. We use MATLAB scripts to control COMSOL to sweep over different particle coordinates. 2-body force tables were obtained from isolated two-particle configurations. Three-particle force tables were obtained from simultaneous three-particle configurations, so that the resulting residual force includes changes in scattering and streaming caused by the full particle arrangement.

The FEM force tables were then used in two ways. First, selected configurations were plotted directly to visualize streaming fields and force vectors around the trimer. Second, the sampled forces were fit to the interpolated spline force model described below, which makes the subsequent stability analysis and MD relaxation computationally feasible. The interpolation step does not assume pairwise additivity or reciprocity; it stores the force on each particle as a function of the full three-particle geometry.

Fig.~\ref{fig:supp-gorkov-slices} shows the no-particle FEM horn calculation used to estimate the large-scale in-plane confinement. Because this calculation contains no levitated particles, its spatial scale is much larger than the particle-scale force simulations.

\textit{Deformation-space projection—}The main text defines the deformation coordinates $\Delta=(\mathrm{E},\mathrm{PS}_{1},\mathrm{PS}_{2})$ and the decomposition of the particle force into 1-body, 2-body, and 3-body parts. For implementation, the six in-plane particle coordinates are first expanded in an orthonormal basis consisting of translations $T_x,T_y$, global rotation $R$, isotropic expansion $E$, and the two pure-shear modes $\mathrm{PS}_{1},\mathrm{PS}_{2}$. Only the last three modes are used for the deformation-space force field.

Let
$
    \mathbf{u}
    =
    \left(u_{1x},u_{1y},u_{2x},u_{2y},u_{3x},u_{3y}\right)^{T}.
$
If $\mathbf{e}_\mathrm{E}$, $\mathbf{e}_{\mathrm{PS}_{1}}$, and $\mathbf{e}_{\mathrm{PS}_{2}}$ are the corresponding deformation basis vectors, any six-component particle force field $\mathbf{F}$ is projected as
$
    \mathbf{F}_{\Delta}^{\mathrm{def}}
    =
    \left(
    \mathbf{F}\cdot\mathbf{e}_\mathrm{E},
    \mathbf{F}\cdot\mathbf{e}_{\mathrm{PS}_{1}},
    \mathbf{F}\cdot\mathbf{e}_{\mathrm{PS}_{2}}
    \right).
    \label{eq:supp-projection}
$
This same projection is applied separately to the 1-body, 2-body, and residual 3-body forces defined in the main text, giving $\mathbf{F}_{\Delta}^{(1)}$, $\mathbf{F}_{\Delta}^{(2)}$, and $\mathbf{F}_{\Delta}^{(3)}$. The resulting field $\mathbf{F}_{\Delta}^{(3)}(\mathrm{E},\mathrm{PS}_{1},\mathrm{PS}_{2})$ is the input to the Helmholtz projection below.

\textit{Helmholtz projection—}
In the three-dimensional deformation space, the 3-body force is decomposed as
$
    \mathbf{F}_{\Delta}^{(3)}
    =
    \mathbf{F}_{\Delta}^{(3,\mathrm{cons})}
    +
    \mathbf{F}_{\Delta}^{(3,\mathrm{nc})}
    =
    \nabla_{\Delta}\phi
    +
    \nabla_{\Delta}\times\mathbf{A}.
    \label{eq:supp-helmholtz}
$
The conservative component is the curl-free projection of the field and is represented by the scalar potential $\phi$. The non-conservative component is the divergence-free remainder and is represented by the vector potential $\mathbf{A}$.

Numerically, the force field is sampled on a regular grid in $(\mathrm{E},\mathrm{PS}_{1},\mathrm{PS}_{2})$. Before the decomposition, the 3-body force is averaged over particle relabelings so that the interpolated field respects the permutation symmetry of the identical particles. Only the residual 3-body force is decomposed into conservative and non-conservative components. The 1-body confinement and 2-body force are not included in the Helmholtz solve, because the acoustic confinement and isolated pair interaction are conservative by construction; they are added to the conservative channel after the 3-body residual has been decomposed.

The sign of the scalar potential is conventional. In the numerical implementation we write $\varphi=-\phi$, so that
$
    \mathbf{F}_{\Delta}^{(3,\mathrm{cons})}
    =
    -\nabla_{\Delta}\varphi .
$
The potential $\varphi$ is obtained from the finite-difference Poisson problem
$
    \nabla_{\Delta}^{2}\varphi
    =
    -\nabla_{\Delta}\cdot\mathbf{F}_{\Delta}^{(3)}
    \label{eq:supp-poisson}
$
using the same discrete derivative operators used for the divergence and curl diagnostics. The sparse linear system is solved by least squares, which fixes the curl-free component whose divergence matches the divergence of the input field on the finite computational box. The conservative component is then
$
    \mathbf{F}_{\Delta}^{(3,\mathrm{cons})}
    =
    -\nabla_{\Delta}\varphi,
$
and the non-conservative component is the residual
$
    \mathbf{F}_{\Delta}^{(3,\mathrm{nc})}
    =
    \mathbf{F}_{\Delta}^{(3)}
    -
    \mathbf{F}_{\Delta}^{(3,\mathrm{cons})}.
$
This construction ensures that the conservative component has zero curl up to discretization error, while the non-conservative component contains the part of the force field that cannot be represented by a scalar potential in the deformation coordinates. The 3-body field and the scalar potential are additionally averaged over the threefold rotations of the equilateral reference state, and the potential is mildly smoothed before reconstructing $\mathbf{F}_{\Delta}^{(3,\mathrm{cons})}$. The non-conservative component is then recomputed by residual subtraction so that the conservative and non-conservative parts exactly assemble back to the symmetrized 3-body field. The stiffness contributions shown in the main text are obtained by projecting each component onto the least-stable pure-shear direction at the equilateral configuration.

\textit{Interpolated spline force model—}The simulated 3-body residual is represented as Radial Basis Functions (RBFs), functions of the local geometry around each particle. For a particle $i$ in a triplet $(i,j,k)$, the input coordinates are
$
    \mathbf{x}_{i,jk}
    =
    \left(
    r_{ij},
    r_{ik},
    \theta_{jik}/180^\circ
    \right),
$
where $r_{ij}=|\mathbf{r}_{j}-\mathbf{r}_{i}|$, $r_{ik}=|\mathbf{r}_{k}-\mathbf{r}_{i}|$, and $\theta_{jik}$ is the angle between $\mathbf{r}_{j}-\mathbf{r}_{i}$ and $\mathbf{r}_{k}-\mathbf{r}_{i}$. Distances are measured in particle diameters. The 3-body force on particle $i$ is stored in an orthogonal local basis,
$
    \mathbf{F}_{i,jk}^{(3)}
    =
    \left[
    f_{\mathrm{rad}}(\mathbf{x}_{i,jk})\,\hat{\mathbf{r}}_{ij}
    +
    f_{\mathrm{tan}}(\mathbf{x}_{i,jk})\,\mathbf{e}_{\perp}
    \right],
    \label{eq:supp-orthogonal-force}
$
The radial basis vector is $\hat{\mathbf{r}}_{ij}=(\mathbf{r}_{j}-\mathbf{r}_{i})/r_{ij}$, and
$
    \mathbf{e}_{\perp}
    =
    \mathrm{sgn}\!\left[
    (\mathbf{r}_{j}-\mathbf{r}_{i})\times
    (\mathbf{r}_{k}-\mathbf{r}_{i})
    \right]_{z}
    \left(-\hat{r}_{ij,y},\hat{r}_{ij,x}\right)
$
fixes the tangential direction using the orientation of the triplet.

The RBF fit used a thin-plate-spline kernel, $200$ neighbors, smoothing $10^{-6}$, and a linear polynomial tail. This fit was then converted into an interpolated spline force model for fast force evaluation. The spline was evaluated on a $400\times 400\times 180$ grid in $(r_{ij},r_{ik},\theta/180^\circ)$ over $r_{ij},r_{ik}\in[1,5]$ and $\theta/180^\circ\in[10^{-6},1-10^{-6}]$. The two coefficient arrays were stored as single-precision grids and evaluated with trilinear interpolation.

\textit{Molecular-dynamics relaxation—}The MD calculations were deterministic Langevin relaxations with a very small nominal temperature. We integrated unit-mass equations of motion with time step $\Delta t=0.001$, damping time $T_{\mathrm{damp}}=30$, and noise scale $T=10^{-10}$. The damping factor was retained even though the noise amplitude was negligible, so the trajectories relax to stable driven fixed points rather than sampling a thermal ensemble. The confinement sweep used $50$ values of $k$ between $10^{-6}$ and $10^{-3}$, with extra points in the transition window from $3\times10^{-5}$ to $3\times10^{-4}$. Each run used $10^7$ integration steps and saved one frame every $10^4$ steps.

\textit{AI-assisted tool usage-}OpenAI Codex GPT-5.5 High was used under author direction to assist with code organization and debugging, generation and revision of analysis and plotting scripts, exploration of diagnostic analyses, and limited editing of explanatory text for clarity. The AI tools were not used as authors and did not replace the authors’ scientific judgment. All simulations, numerical analyses, figures, and conclusions were directed, reviewed, and verified by the authors through inspection of the underlying code, raw and processed data, regenerated outputs, and/or consistency checks against the experimental observations. The authors take full responsibility for the accuracy, originality, and interpretation of all content in the manuscript and End Matter.

\end{document}